\documentclass[a4paper]{article}

\usepackage{multirow} %my%
\usepackage{cite} %my%
\usepackage{subfigure} %my%
\usepackage{bbm} %my%

\usepackage{INTERSPEECH2021}

\title{Fine-grained Style Modeling, Transfer and Prediction in Text-to-Speech Synthesis via Phone-Level Content-Style Disentanglement}

\name{Daxin Tan, Tan Lee}
\address{Department of Electronic Engineering, The Chinese University of Hong Kong, Hong Kong}
\email{daxintan@link.cuhk.edu.hk, tanlee@ee.cuhk.edu.hk}

\begin{document}

\maketitle

\begin{abstract}
\vspace{-0.5em}
This paper presents a novel design of neural network system for fine-grained style modeling, transfer and prediction in expressive text-to-speech (TTS) synthesis. Fine-grained modeling is realized by extracting style embeddings from the mel-spectrograms of phone-level speech segments. Collaborative learning and adversarial learning strategies are applied in order to achieve effective disentanglement of content and style factors in speech and alleviate the ``content leakage'' problem in style modeling. The proposed system can be used for varying-content speech style transfer in the single-speaker scenario. The results of objective and subjective evaluation show that our system performs better than other fine-grained speech style transfer models, especially in the aspect of content preservation. By incorporating a style predictor, the proposed system can also be used for text-to-speech synthesis. Audio samples are provided for system demonstration\footnote{https://daxintan-cuhk.github.io/pl-csd-speech}.
\end{abstract}


\vspace{-1em}
\section{Introduction}
\vspace{-0.5em}
Human speech production manifests a complex integration of physical, cognitive and affective processes. The realization of a spoken utterance involves three main factors, namely the content factor, the speaker factor and the style factor. The content factor is determined by the linguistic content of speech. The speaker factor refers to voice characteristics that are pertinent to recognizing the speaker. While an unambiguous definition of ``style'' may not exist, in this study the style factor is assumed to cover broadly any aspect of the speech utterance that is not determined by its linguistic content and the speaker's inherent voice characteristics \cite{liu2020expressive}. The style of speech is often related to speaking situation, attitude, emotion, language proficiency, etc.

Different tasks of spoken language processing can be viewed as purposeful processes of analyzing and manipulating one or more of the three factors. Automatic speech recognition (ASR)\cite{gulati2020conformer,han2020contextnet} and text-to-speech synthesis (TTS)\cite{wang2017tacotron, shen2018natural,li2019neural,ren2019fastspeech,ren2020fastspeech,yu2019durian} are focused on the content factor, with the other two factors being suppressed or ignored. Speaker recognition\cite{watanabe2018espnet,snyder2018x} and voice conversion\cite{qian2019autovc,chou2019one,kaneko2019stargan} aim to capture and manipulate the speaker factor, while emotion recognition\cite{akccay2020speech} and expressive speech synthesis deal primarily with the style factor. 
Relating to expressive TTS, speech style transfer (SST)\cite{skerry2018towards, wang2018style, zhang2019learning, kenter2019chive, hsu2018hierarchical, lee2019robust, klimkov2019fine, hu2020unsupervised, karlapati2020copycat, li2021controllable, zhang2020learning} refers to the process of transferring the style of one speech utterance (reference speech) into another utterance (source speech). In a neural network based SST model, style-related embedding is obtained from the reference speech, and subsequently combined with the text embedding of source speech (and probably a speaker embedding) to condition the generation of output speech.

Style embedding can be defined and extracted at different levels of granularity. The studies in \cite{skerry2018towards} and \cite{wang2018style} extended the Tacotron system with the use of fix-length style embedding at utterance level. Variational auto-encoder (VAE) and hierarchical structure were applied to improve the representation capability of learned style embeddings in \cite{zhang2019learning}, \cite{kenter2019chive} and \cite{hsu2018hierarchical}. To facilitate fine-grained style control on specific parts of an utterance, phone-level style embedding was investigated in \cite{lee2019robust} and \cite{klimkov2019fine}. In \cite{lee2019robust}, a secondary attention module was used to generate style embedding from mel-spectrogram of reference speech. In \cite{klimkov2019fine}, style embedding was derived from pitch and intensity features aggregated at phone level.

The main issue in the extraction of style embedding is ``content leakage''\cite{hu2020unsupervised}. Typically, an SST model comprises a style encoder and a text encoder. The style encoder aims to encode the style factor of reference speech into a style embedding, while the text encoder generates a text embedding from the text of source speech. The entire model is trained in an end-to-end manner with the goal of reconstructing the input utterance. The same training utterance acts as both reference and source speech, i.e., its acoustic features go to the style encoder and its transcription goes to the text encoder. Since no constraint is exerted to enforce the style encoder to focus exclusively on style-related information of input speech, a certain degree of content information may be ``leaked'' into the learned style embedding. In the worst case, the text encoder does not take any effect such that the the style embedding contains all information required for reconstructing the input speech. Such ``corrupted'' style embedding would not be useful for style transfer when the source speech and reference speech is different in content. In \cite{hu2020unsupervised}, this problem was addressed by minimizing the mutual information between content and style embedding during training. In \cite{ma2018neural}, a pairwise training strategy was proposed to enforce correct mapping from input text to different speech utterances.

This paper presents a novel design of neural network model for fine-grained style modeling, transfer and prediction in expressive text-to-speech (TTS) synthesis. For fine-grained modeling and control, style embeddings are generated directly from mel-spectrogram at phone level. In order to achieve effective disentanglement of content and style factors, we propose to apply collaborative learning to force the content encoder to focus on phone identification and apply adversarial learning to force the style encoder to ignore content information. With a properly trained style predictor, the proposed system can be regarded as a unified framework for both speech style transfer and text-to-speech synthesis without reference speech.

\begin{figure*}[h]
  \centering
  \includegraphics[width=\linewidth, trim=60 360 100 45]{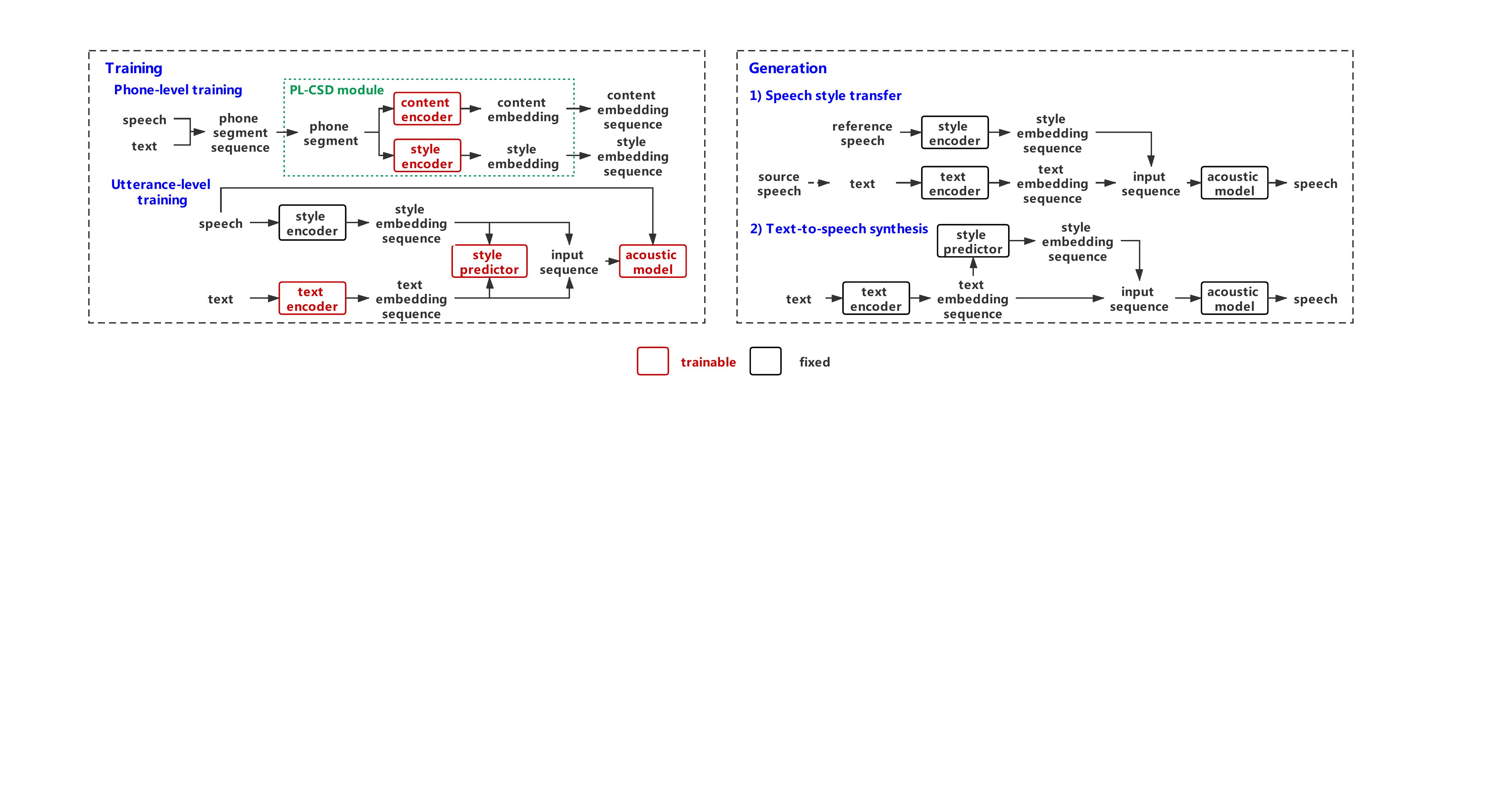}
  \caption{Overview of the proposed system}
  \label{fig:model overview}
\vspace{-1em}
\end{figure*}

\vspace{-1em}
\section{The Proposed System}
\vspace{-0.5em}
The proposed system is depicted as in Figure \ref{fig:model overview}. The training process comprises two parts: phone-level training and utterance-level training. The phone-level content-style disentanglement (PL-CSD) module is obtained by phone-level training, and subsequently used in utterance-level training. Upon completion of model training, the system can be used for speech style transfer and text-to-speech synthesis.

\vspace{-0.5em}
\subsection{Phone-level training}
\vspace{-0.5em}
The PL-CSD module is trained at this stage. Let $u_i$ denote the $i^{th}$ utterance and $t_i$ be the corresponding text transcription. $t_i$ is expressed as a phone sequence, i.e., $t_i=[ w_1,..., w_{m_i} ]$, where $m_i$ is the number of phones, and $w_k$ denotes the $k^{th}$ phone. By applying forced alignment, with silence/pause segments excluded, $u_i$ is divided into $m_i$ segments, denoted as $\{s_1,..., s_{m_i}\}$, where segment $s_k$ corresponds to phone $w_k$. Without considering the temporal dependency, the collection of $(s_k, w_k)$ are treated as independent data instances for training PL-CSD. $s_k$ is represented by frame-level mel-spectrogram features $[f_1,...,f_{n_k}]$, where $n_k$ is the number of frames in $s_k$. Note that $w_k$ must be one of the phones in the concerned language. In this study, we use the $39$ English phones as defined in the APARBET.  

\begin{figure}[h]
\vspace{-1em}
  \centering
  \includegraphics[width=\linewidth, trim=30 450 50 30]{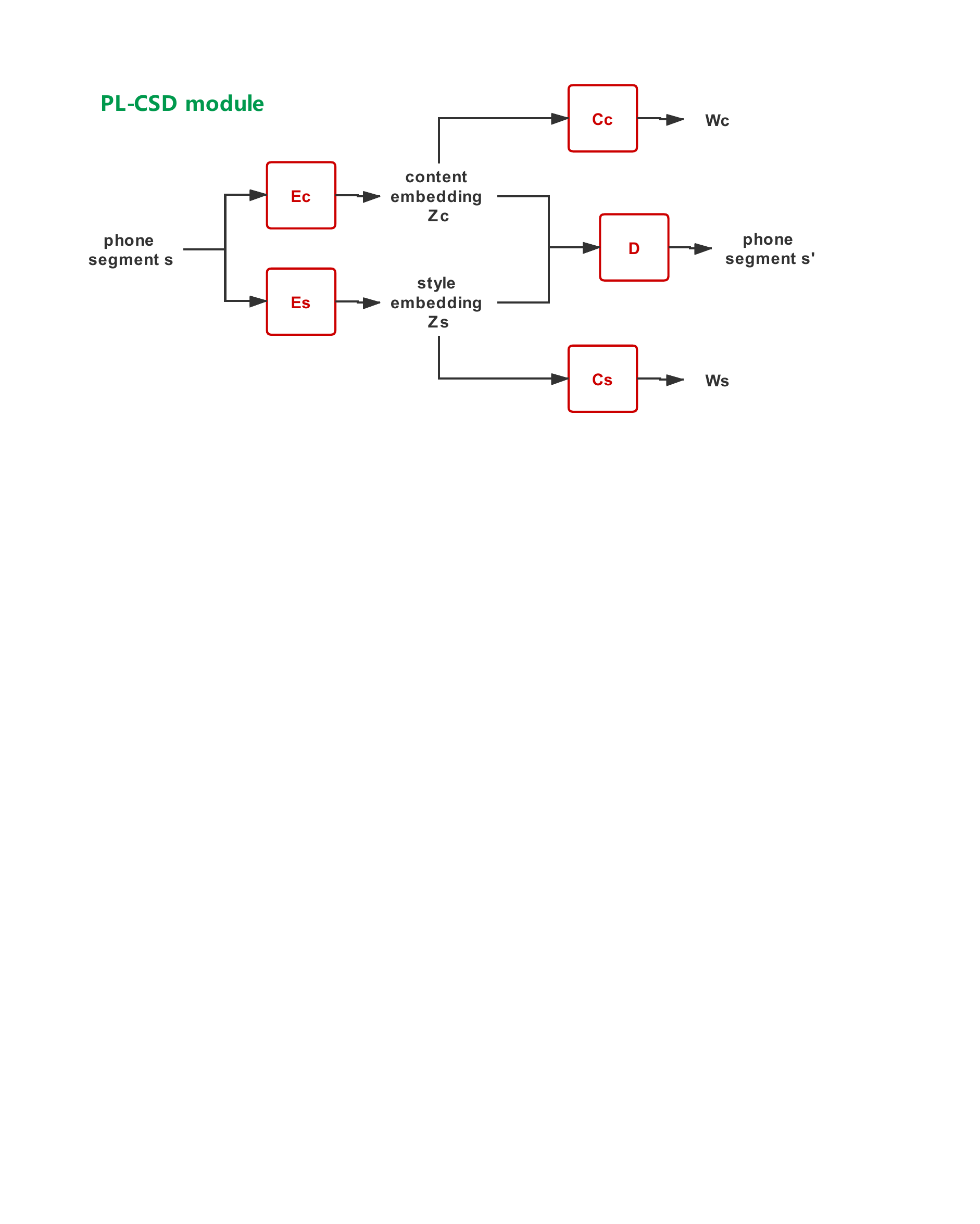}
  \caption{Detailed design of the PL-CSD module}
  \label{fig:PL-CSD_module}
\vspace{-1em}
\end{figure}

As shown in Figure\ref{fig:PL-CSD_module}, the PL-CSD module consists of the following components:

\noindent\textbf{Content encoder ($E_c$):} to encode the content factor of phone segment $s$ into the content embedding $z_c$, i.e., $z_c=E_c(s)$ 

\noindent\textbf{Style encoder ($E_s$):} to encode the style factor of phone segment $s$ into the style embedding $z_s$, i.e., $z_s=E_s(s)$

\noindent\textbf{Content-to-phone classifier ($C_c$):} to predict the phone identity $w_c$ from the content embedding $z_c$, i.e., $w_c=C_c(z_c)$ 

\noindent\textbf{Style-to-phone classifier ($C_s$):} to predict the phone identity $w_s$ from the style embedding $z_s$, i.e., $w_s=C_s(z_s)$ 

\noindent\textbf{Decoder ($D$):} to reconstruct phone segment $s'$ from the content embedding $z_c$ and the style embedding $z_s$, i.e., $s'=D(z_c,z_s)$

\noindent\textbf{Segment classifier ($C_{seg}$):} to discriminate if a phone segment is from natural speech (true) or synthesized (false), i.e., $P_{true}=C_{seg}(\{s,s'\})$)

\vspace{-0.5em}
\subsubsection{Auto-encoder training}
\vspace{-0.5em}
The content encoder $E_c$, the style encoder $E_s$ and the decoder $D$ together make up an auto-encoder model. The difference between the reconstructed segment and the original one is measured by the $L2$ norm of their mel-spectrograms, which is referred to as the spectrogram loss $L_s$, and the binary cross-entropy loss of the gates $L_g$, which indicates the end of segment \cite{shen2018natural}. For decoder training, ground-truth mel-spectrogram of the current frame is concatenated with the content embedding and the style embedding to predict the mel-spectrogram of the next frame. The overall auto-encoder loss $L_{auto}$ is given as,
\begin{equation}
L_{auto}(\theta_{E_c},\theta_{E_s},\theta_{D})=\sum\limits_{s_k}\sum\limits_{L_s, L_g}L(s_k, D(E_c(s_k), E_s(s_k)))
\vspace{-1em}
\end{equation}
where $\theta$ refers to the model parameters to be trained. 

Basic auto-encoder training is not expected to achieve disentanglement, as the roles of the content encoder and style encoder are not differentiated. As described below, collaborative training is suggested for training the content encoder while adversarial training is applied to the style encoder.

\vspace{-0.5em}
\subsubsection{Collaborative training of content encoder}
\vspace{-0.5em}
The content embedding is expected to carry pertinent information to phone identification. An auxiliary content-to-phone classifier is introduced to assess the goodness of content embedding. By training the content encoder and the content-to-phone classifier collaboratively, the content embedding is forced to capture phone identity information. The content-to-phone classification loss is defined as,
\begin{equation}
L_{c}(\theta_{E_c},\theta_{C_c})=\sum\limits_{(s_k,w_k)} -\log P(w_k|C_c(E_c(s_k))).
\vspace{-0.5em}
\end{equation}
To ensure that content embeddings extracted from different segments that belongs to the same phone are similar, the following contrast loss $L_{contra}$ is imposed,
\begin{equation}
L_{contra}(\theta_{E_c})=\sum\limits_c \mathbbm{1}_{w_i=w_j} ||E_c(s_i)-E_c(s_j)||_2
\vspace{-1em}
\end{equation}

\vspace{-0.5em}
\subsubsection{Adversarial training of style encoder}
\vspace{-0.5em}
In contrast to the content embedding, the style embedding is desired not to encode any information about phone identity, or that the phone carried by a segment should be non-identifiable from its style embedding. This is achieved via adversarial training \cite{goodfellow2014generative}. In the discrimination phase, the parameters of style encoder are fixed, and the style-to-phone classifier is trained to perform phone identification from the style embedding,
\begin{equation}
L_{s}^{dis}(\theta_{C_s})=\sum\limits_{(s_k,w_k)} -\log P(w_k|C_s(E_s(s_k)))
\vspace{-0.5em}
\end{equation}
In the generation phase, the parameters of style-to-phone classifier are fixed, and the style encoder is trained such that the segment's phone identity cannot be predicted from the style embedding. The following loss is defined to enforce equal posterior probabilities across all phones,
\begin{equation}
L_{s}^{gen}(\theta_{E_s})=\sum\limits_{s_k}\sum\limits_{w \in N_w}||P(w|C_s(E_s(s_k)))-\frac{1}{N_w}||_2
\vspace{-0.5em}
\end{equation}
where $N_w$ denotes the total number of phones.

\vspace{-0.5em}
\subsubsection{Adversarial enhancement}
\vspace{-0.5em}
An adversarial enhancement process is applied to supplement the model's reconstruction function. A discriminator network $C_{seg}$ is utilized to judge if a given segment is from natural speech or synthesized, while the auto-encoder serves as a generator model to confuse the discriminator. For the discriminator, we have, $s_k'=D(E_c(s_k), E_s(s_k))$,
\begin{equation}
L_{seg}^{dis}(\theta_{C_{seg}})=\sum\limits_{s_k}-[\log C_{seg}(s_k)+\log(1-C_{seg}(s_k')]
\vspace{-0.5em}
\end{equation}
For the generator, the loss is given as,
\begin{equation}
L_{seg}^{gen}(\theta_{E_c},\theta_{E_s},\theta_{D})=\sum\limits_{s_k}-\log C_{seg}(s_k')
\vspace{-1em}
\end{equation}

\vspace{-0.5em}
\subsubsection{Overall training algorithm}
\vspace{-0.5em}
The overall training algorithm for PL-CSD is in Table \ref{tab:algorithm}.

\begin{table}[h]
\vspace{-0.5em}
\caption{PL-CSD training algorithm}
\vspace{-0.5em}
\label{tab:algorithm}
\begin{tabular}{l}
\hline
\textbf{Input:} segments and corresponding phones: $(s_k, w_k)$\\
\hline
\textbf{Repeat until convergence:}\\
1. Train $E_c$, $E_s$, $D$ by minimizing Eq. (1)\\
2. Train $E_c$, $C_c$ by minimizing Eq. (2) and Eq. (3)\\
3. Fix $E_s$, train $C_s$ by minimizing Eq. (4)\\
4. Fix $C_s$, train $E_s$ by minimizing Eq. (5)\\
5. Fix $E_c$, $E_s$, $D$, train $C_{seg}$ by minimizing Eq. (6)\\
6. Fix $C_{seg}$, train $E_c$, $E_s$, $D$ by minimizing Eq. (7)\\
\hline
\end{tabular}
\vspace{-1em}
\end{table}

\vspace{-0.5em}
\subsection{Utterance-level training}
\vspace{-0.5em}
Upon completion of phone-level training, utterance-level training is carried out to optimize the text encoder and the acoustic model for speech generation. The process is similar to the training of a basic neural TTS system. For each training utterance with text transcription, a sequence of phone-level style embeddings are obtained from the style encoder. The text transcription is converted to a phone sequence, which is passed to the text encoder to generate the text embedding sequence. The style embedding sequence and text embedding sequence are of the same length. They are combined by concatenating the two embeddings for each phone in the sequence. The acoustic model takes in the combined embedding sequence and outputs the mel-spectrogram of speech. The text encoder and the acoustic model are optimized jointly to minimize the mean square error between the generated mel-spectrogram and the ground-truth one.

After this training process, the derived text embedding sequence and the style embedding sequence for each utterance serve as the training pair of style predictor. The basic function of the style predictor is to map a text embedding sequence to a style embedding sequence.

\vspace{-0.5em}
\subsection{Generation}
\vspace{-0.5em}
Upon completion of phone-level and utterance-level training as described above, the system is ready for speech generation from given text. Two cases of speech generation are supported: speech style transfer and text-to-speech synthesis. In the speech style transfer case, the style encoder processes the reference speech to derive the style embedding sequence. The text encoder processes the phone sequence that converted from the input text (corresponding to the source speech) and generates the text embedding sequence. As the two embedding sequences may have different lengths, linear interpolation is carried out on the style embedding sequence, in order to make it the same length as the text embedding sequence. These two embedding sequences are concatenated at each time step, The combined sequence is then presented to the acoustic model for speech generation. In the case of text-to-speech synthesis, the input text is first converted into phone sequence. The style predictor is used to derive the style embedding sequence from the phone sequence, while the text encoder generates the text embedding sequence as usual. The two embedding sequences are combined and presented to the acoustic model for speech generation.

\vspace{-0.5em}
\section{Experimental Setup}
\vspace{-0.5em}
The LJ speech dataset \cite{ljspeech17} is used in this study. It contains $13,100$ utterances from a single speaker. Their total length is about $24$ hours. $90\%$ of the utterances are used as training data and the remaining $10\%$ as test data. Forced alignment of speech is carried out using the Montreal Forced Aligner\cite{mcauliffe2017montreal} based on the CMU Pronouncing Dictionary. Mel-spectrogram are obtained with a similar setting to the standard Tacotron 2 model. In the PL-CSD module, both the content encoder and the style encoder are made up by bidirectional LSTM of dimension $512$ ($256$ in each direction), in which the last cell state is projected to embedding via a linear layer. The decoder contains a unidirectional LSTM of dimension $512$, with two linear layers applied on the output at each time step. One of them predicts the mel-spectrogram and the other indicates the end of sequence. The dimensions of content embedding and style embedding are both $64$. For utterance-level training, the acoustic models adopt the standard Tacotron 2 structure, where the dimension of text embedding is $64$. The WaveGlow vocoder\cite{prenger2019waveglow} is used to generate speech waveform from the predicted mel-spectrogram. The style predictor adopts the feed-forward Transformer block structure as in \cite{ren2019fastspeech} and \cite{ren2020fastspeech}, which is a stack of self-attention layer and 1D-convolution.

\vspace{-0.5em}
\section{Results and Discussion}

\subsection{Visualization of embeddings}
\begin{figure}[h]
\centering
\subfigure[Vowel content embeddings]{
\includegraphics[width=0.43\linewidth,trim=50 50 50 50 ]{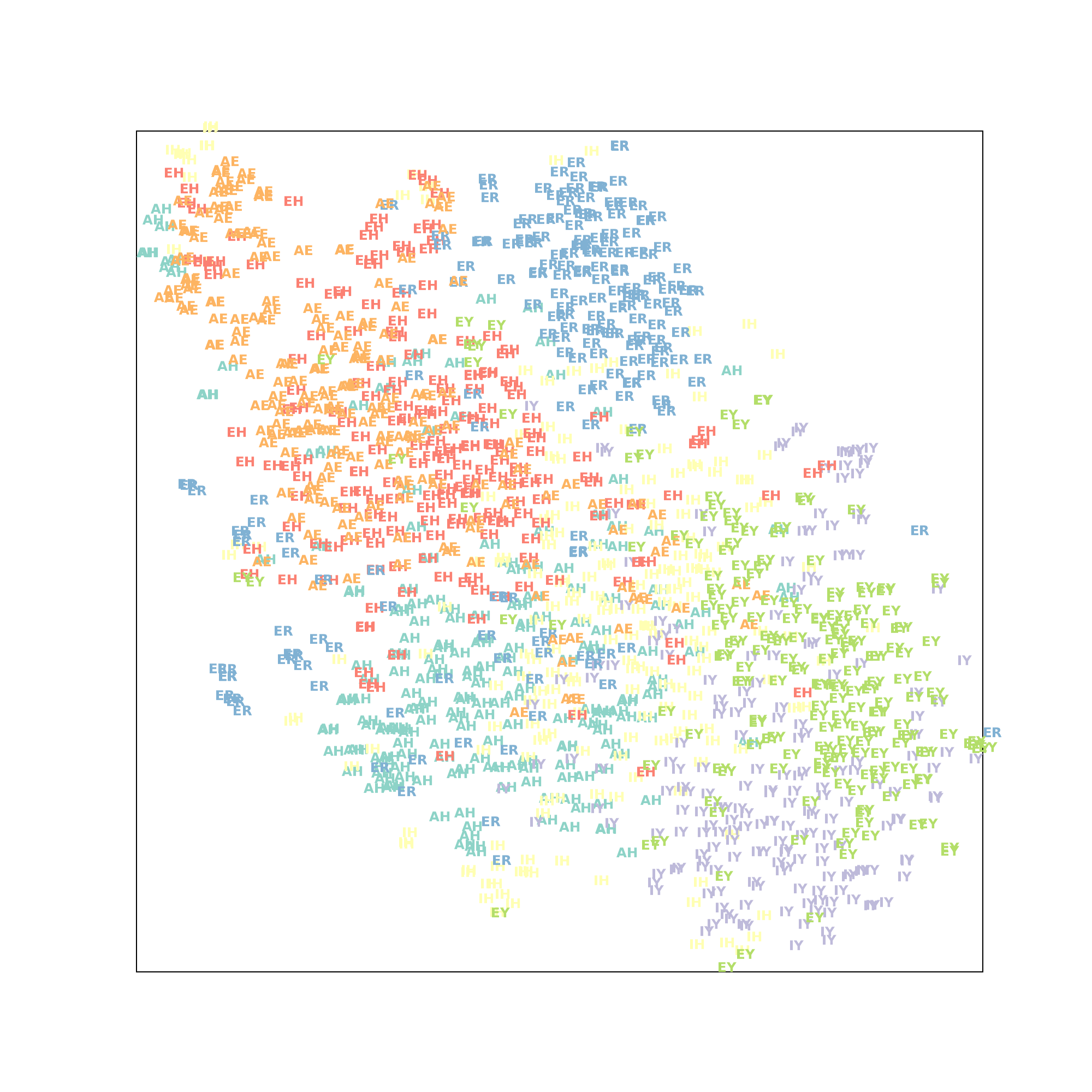}
}
\hspace{0.02\linewidth}
\subfigure[Vowel style embeddings]{
\includegraphics[width=0.43\linewidth, trim=50 50 50 50]{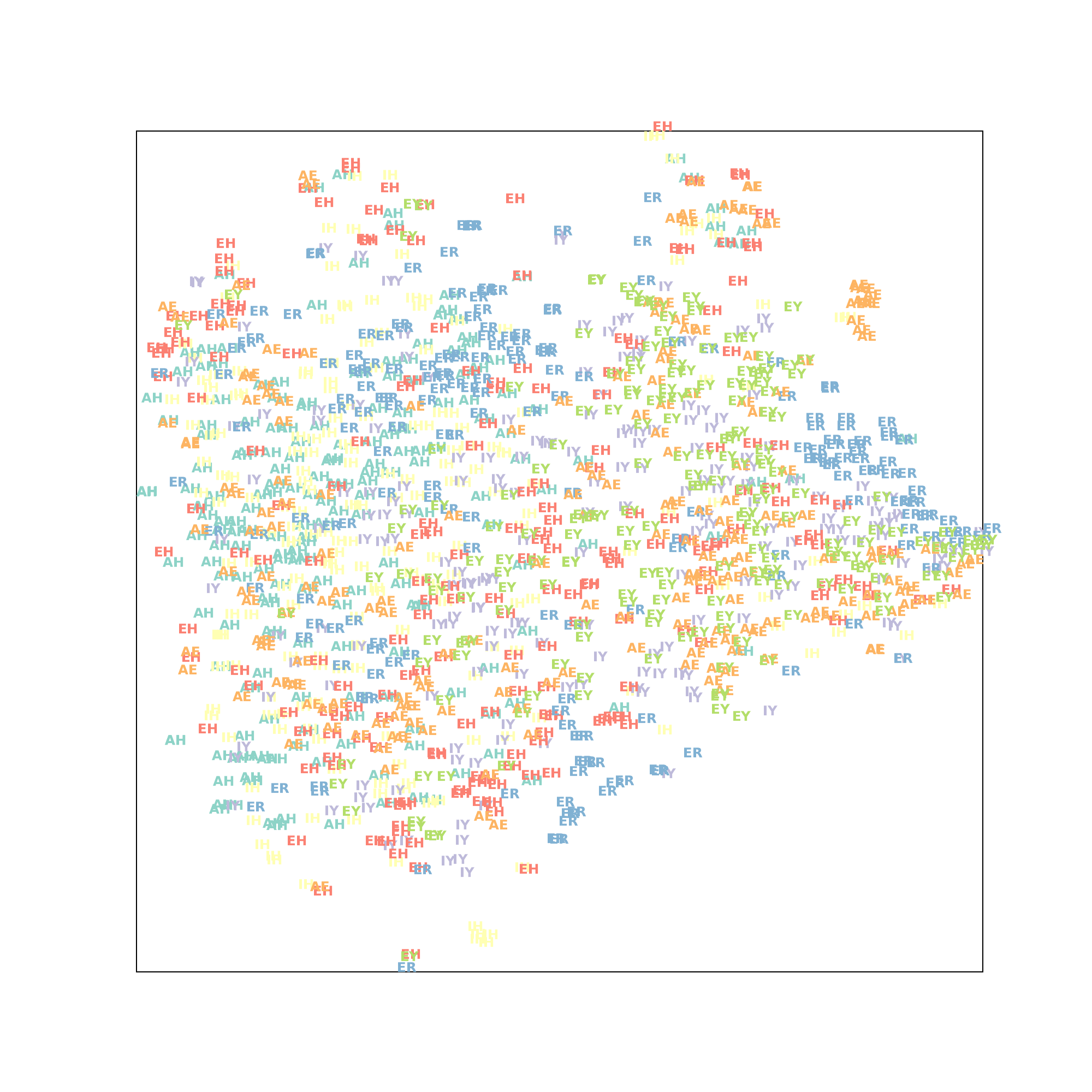}
}
\vspace{-0.5em}
\caption{The t-SNE visualization of content embeddings and style embeddings for vowel segments.}
\label{fig:tsne_embedding}
\vspace{-1em}
\end{figure}

\vspace{-0.5em}
T-distributed Stochastic Neighbor Embedding (t-SNE) \cite{maaten2008visualizing} is applied for visualization. The analysis is carried out with $7$ English vowels that occur most frequently in the dataset. For each phone, $200$ segments are randomly selected from the test set, and their content and style embeddings are derived from the trained content and style encoder respectively. Figure \ref{fig:tsne_embedding} shows the t-SNE scatter plots of the content embeddings and the style embeddings of the vowel segments, where each color represents a kind of vowel. It can be seen that the content embeddings of segments of same kind of vowel tend to fall into certain clusters. On the other hand, the style embeddings of different vowel segments do not seem to be separable. This indicates that the phone identity information can be well captured by content embedding, but not present in style embedding.

\renewcommand{\thesubfigure}{} % change the subfigure title

\vspace{-0.5em}
\subsection{Evaluation of generated speech}
\vspace{-0.5em}
Two SST systems in \cite{lee2019robust} and \cite{klimkov2019fine} are compared with our system. All of these systems aim at fine-grained style modeling. The system in \cite{lee2019robust} adopted a secondary attention mechanism to extract the style embedding implicitly while our system carries out extraction with explicit content-style disentanglement. The system in \cite{klimkov2019fine} derived style embedding on pitch and intensity while our system uses mel-spectrogram as input features.

\vspace{-0.5em}
\subsubsection{Objective evaluation}
\vspace{-0.5em}
Being able to reconstruct an input utterance from its text embedding and style embedding is a basic requirement to guarantee no information is lost in the disentanglement process. In this part of evaluation, the source speech and the reference speech are from the same utterance in the test set. The similarity between the synthesized speech and the original speech is used to measure the performance of reconstruction. Following \cite{skerry2018towards}, we use Voicing Decision Error (VDE), Gross Pitch Error (GPE), F0 Frame Error (FFE) and Mel Cepstral Distortion (MCD) as the evaluation metrics. As shown in Table \ref{tab:reconstruction}, the proposed model shows slightly better or comparable performance than the two previously reported systems.

\begin{table}[h]
\vspace{-0.5em}
\caption{Objective evaluation on speech reconstruction}
\vspace{-0.5em}
\label{tab:reconstruction}
\begin{small}
\centering
\begin{tabular}{c|c|c|c|c}
\hline
 System &  VDE(\%)$\downarrow$ & GPE(\%)$\downarrow$ & FFE(\%)$\downarrow$ & MCD$\downarrow$ \\
\hline
% second attention
Lee\cite{lee2019robust} & \textbf{9.05} & 8.16 & \textbf{13.04} & 10.91\\
\hline
% Amazon
Klimkov\cite{klimkov2019fine} &  12.00 & 5.43 & 14.53 & 11.67\\
\hline
Ours &  11.03 & \textbf{4.57} & 13.15 & \textbf{10.49} \\
\hline
\end{tabular}
\end{small}
\vspace{-1em}
\end{table}

When the proposed system is used for style transfer, the reference speech may have different text content from the source speech. If the content-style disentanglement is successful, the synthesized speech is expected to contain the same content as the source speech and carry the style of reference speech. We use an end-to-end ASR system provided in the ESPnet toolkit \cite{watanabe2018espnet} to evaluate the content similarity of synthesized speech and source speech. The word error rate (WER) and phone error rate (PER) on synthesized speech are evaluated with respect to the transcription of source speech. The system in \cite{lee2019robust} is found to have serious ``content leakage'' problem and fails to retain the source content, i.e., the synthesized speech carries both the content and the style from the reference speech and neglect the source speech. Compared with the system of \cite{klimkov2019fine}, our system is significantly better in preserving the source content. The comparison is shown as in Table \ref{tab:recombination}.

\begin{table}[h]
\vspace{-0.5em}
\caption{ASR error rates on style-changed speech}
\vspace{-0.5em}
\label{tab:recombination}
\centering
\begin{tabular}{c|c|c}
\hline
System & WER(\%)$\downarrow$ & PER(\%)$\downarrow$ \\
\hline
% second attention
Lee\cite{lee2019robust} & 90.4 & 74.1 \\
\hline
% amazon model
Klimkov\cite{klimkov2019fine} &  29.2  & 14.5\\
\hline
Ours & \textbf{21.4}  & \textbf{8.5} \\
\hline
\end{tabular}
\vspace{-1em}
\end{table}

\vspace{-0.5em}
\subsubsection{Subjective evaluation}
\vspace{-0.5em}
Subjective listening tests were carried out to evaluate reconstructed speech and style-transferred speech separately. For evaluating speech reconstruction, the listeners are required to rate the overall similarity between synthesized speech and source speech (natural speech), For evaluating style transfer, separate ratings are given on the content similarity between synthesized speech and source speech, as well as the style similarity between synthesized speech and reference speech. The score of rating ranges from $0$ (completely different) to 5 (exactly the same). A total of 30 listeners participated in the listening test, which was administrated via the Amazon Mechanical Turk platform. Each listener was required to evaluate $30$ sets of test utterances in both cases.

\begin{table}[h]
\vspace{-0.5em}
\caption{MOS results with 95\% confidence intervals}
\vspace{-0.5em}
\label{tab:subjective_evaluation}
\centering
\begin{tabular}{c|c|c|c}
\hline
\multirow{2}*{MOS$\uparrow$} & Reconstruction & \multicolumn{2}{c}{Style transfer}  \\
\cline{2-4}
~ & Overall & Content & Style \\
\hline
% second attention
Lee\cite{lee2019robust} & 2.09±0.17 & - & - \\
\hline
Klimkov\cite{klimkov2019fine} & 2.95±0.13 & 2.97±0.15 &  2.53±0.16 \\
\hline
Ours & \textbf{3.47±0.11} & \textbf{3.41±0.13} & \textbf{2.54±0.15} \\
\hline
\end{tabular}
\vspace{-1em}
\end{table}

Table \ref{tab:subjective_evaluation} shows the results of subjective evaluation. In speech reconstruction, our system attains a significant higher score than baseline systems, which indicates that our style embedding has better representative ability. In speech style transfer, the system in \cite{lee2019robust} is not able to perform varying-content style transfer. Compared with system in \cite{klimkov2019fine}, our system obtained similar rating in the style similarity aspect and significantly higher score in content similarity. This demonstrates that our system performs better in content preservation.

\vspace{-0.5em}
\subsection{Case study}
\begin{figure}[h]
\vspace{-1em}
  \centering
  \includegraphics[width=\linewidth, trim=0 20 0 0]{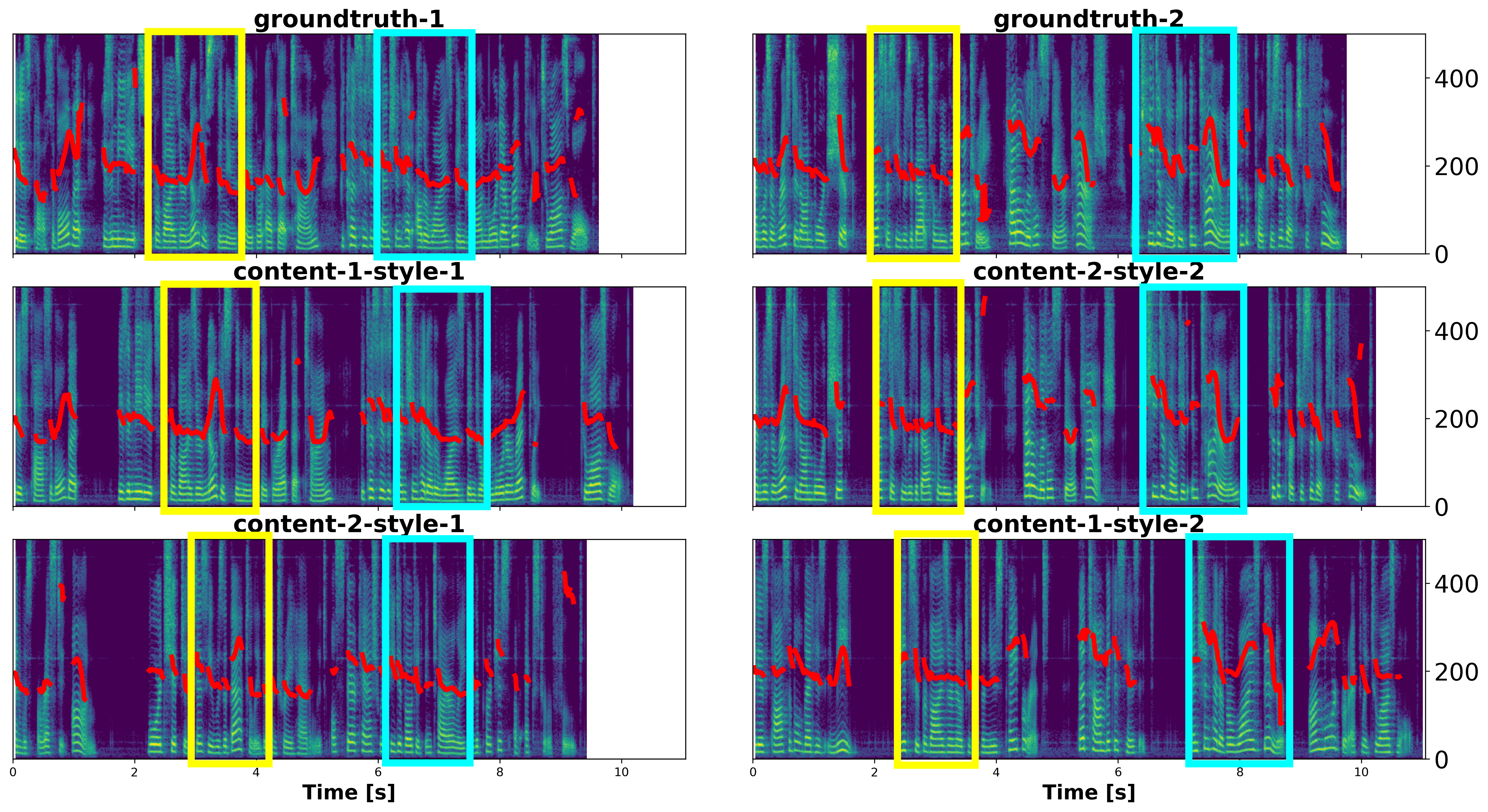}
  \caption{Spectrograms and pitch contours of two test utterances of natural speech and synthesized speech with different combinations of text and style embeddings.}
  \label{fig:speech_example}
\vspace{-1em}
\end{figure}

\vspace{-0.5em}
Figure \ref{fig:speech_example} shows an example of content-style recombination that illustrates the effect of style transfer. The top pane shows the spectrograms and pitch contours of two different utterances of natural speech. The middle pane shows the reconstructed utterances, i.e., synthesized with both text embedding and style embedings coming from the same utterance. The bottom pane shows the results of speech generation with the two utterances' text embeddings (content) swapped. The utterances synthesized with the same style embedding, i.e., those in the same column of the figure, show highly similar patterns of pitch contour. The most notable similarities are marked by the colored boxes.

\vspace{-1em}
\section{Conclusion}

This paper presents a novel system design for fine-grained style modeling, transfer and prediction in expressive text-to-speech synthesis. The proposed collaborative learning and adversarial learning strategies is effective in disentangling style and content factors. The evaluation results show that our system alleviates the ``content leakage'' problem and improves the content preservation in speech style transfer. Further investigation is carried out to extend the system to multi-speaker style transfer.

\vspace{-1em}
\section{Acknowledgement}

This research is partially supported by a Tier 3 funding from ITSP (Ref: ITS/309/18) of the Hong Kong SAR Government, and a Knowledge Transfer Project Fund (Ref: KPF20QEP26) from the Chinese University of Hong Kong.
 
\vfill\pagebreak

\bibliographystyle{IEEEtran}
\bibliography{mybib}

\end{document}